\documentclass[prl,twocolumn,aps,amsmath,amssymb]{revtex4}

\usepackage{graphicx}
\usepackage{dcolumn}
\usepackage{bm}
\usepackage{wrapfig}
\usepackage{amsmath,amsfonts,amssymb,bm,ulem}
\usepackage[usenames]{color}
\usepackage{bm}
\usepackage{slashed}
\usepackage{mathrsfs}
\usepackage{graphicx}
\usepackage{dcolumn}
\usepackage{bm}

\newcommand{\be}{\begin{equation}}
\newcommand{\ee}{\end{equation}}

\newcommand{\scr}{\mathscr}

\begin{document}

\title{Gravitational Decoherence, Asymptotic Quantization, and Entanglement Measures}

\author{Colby DeLisle}
\email[]{cdelisle@phas.ubc.ca}
\author{Jordan Wilson-Gerow}
\author{Philip Stamp}
\affiliation{Department of Physics \& Astronomy \\ and \\ Pacific Institute for Theoretical
Physics\\
University of British Columbia\\ Vancouver, BC V6T 1Z1, Canada}
\date{\today}

\begin{abstract}
Graviton emission decoheres superpositions of matter states. We write the decoherence rate in terms of a ``differential Bondi news" function, and show that this function can be written as an overlap integral between different asymptotic radiative states given by Ashtekar. This result - while derived in the linearized theory - is ultimately expressed for asymptotically flat space-times in terms of variables with unambiguous geometric meaning in the full theory. We treat a simple two-path experiment, and compute the differential news, the purity, the entanglement entropy, and the fidelity, to make explicit the connection with the information loss.
\end{abstract}

\maketitle


Recent years have seen renewed debate over the question of black hole information loss, with the emphasis on the role of soft graviton processes \cite{strominger18,ashtekar18}. Real black holes are hard to analyze, because one needs to treat both the internal black hole quantum states and the horizon; but for a broad class of scattering processes involving soft gravitons, interesting connections have been found between information loss, BMS charges, and gravitational memory \cite{strominger1,pasterski,wilson18,carney}.

Quite generally one can ask how the information in a quantized metric field $g^{\mu\nu}(x)$ flows away from some quantum system ${\cal S}$ with which it is entangled. The most interesting case is where $g^{\mu\nu}(x)$ distinguishes between different internal states of ${\cal S}$; the information flow is then associated with loss of coherence between these internal states. The simplest example is a ``2-path" system, where the internal states are much simpler than those of a black hole, and no horizons are involved. In this case soft gravitons can decohere 2-path interference processes. A number of analyses of such gravitational decoherence have appeared, all in linearized gravity \cite{gravDeco,oniga16,bose17,marletto17}, with sometimes controversial results \cite{anasHu18}; a correct treatment uses gauge invariant radiative states \cite{oniga16,jordanMSc}.

What we show here is that (i) the decoherence functional for the system ${\cal S}$ can be written in terms of the gravitational Bondi news on future null infinity $\scr{I}^+$, and hence in terms of the ``overlap integral'' between Ashtekar's asymptotic coherent states
\cite{Ashtekar1,ashtekar81} peaked at those news configurations; and (ii) one can then analyze simple ``2-path'' experiments, of fundamental interest in quantum gravity, in terms of this overlap integral, as well as using more conventional decoherence measures. We restrict ourselves in this work to situations without horizons, and to matter superpositions which last for a finite time.


\vspace{2mm}

{\bf (i) Decoherence in linearized gravity}: What is the best way to quantify information loss for some system ${\cal S}$? Typically one deals with the reduced density matrix $\rho(1,1') = \langle 1| \hat{\rho}|1' \rangle$ for ${\cal S}$, after tracing over gravitational variables; here $1,1'$ refer to states of $\rho$ created by operators $\mathcal{O}_{j}$, such that $|j\rangle = \mathcal{O}_{j}|0\rangle$. Then $\rho$ propagates according to
\begin{equation}
\rho(2,2')\;=\; \int d1\int d1' \mathcal{K}(2,2';1,1')\; \rho(1,1')
 \label{rho+K}
\end{equation}

Writing variables in pairs (eg., ${\bf 2} \equiv (2,2')$), so that $\rho({\bf 2})\;=\; \int \mathcal{K}({\bf 2};{\bf 1})\; \rho({\bf 1})$, we can write
the propagator ${\cal K}$ in the Feynman-Vernon \cite{feynV63} form
\begin{equation}
\label{eq:dmprop}
\mathcal{K}({\bf 2};{\bf 1})=\int \mathcal{D}{\bf X} \,\mathcal{O}_{\bf 2}\,
\mathcal{O}_{\bf 1} \, e^{i \delta S_M[{\bf X}]}
\mathcal{F}[{\bf X}],
\end{equation}
in which ${\bf X} = (X,X')$ represents the matter variables with action $S_M$, $\delta S_M[{\bf X}] = S_M[X] - S_M[X']$, and $\mathcal{F}[{\bf X}]$ is the famous ``influence functional", incorporating the trace over all gravitational degrees of freedom.

In linearized gravity, with a coupling ${\cal L}_{int} = -\tfrac{1}{2} \int d^4 x h^{\mu\nu} T_{\mu\nu}$ between the stress-energy $T_{\mu\nu}$ and the graviton field $h^{\mu\nu} = g^{\mu\nu} - \eta^{\mu\nu}$, one can find an explicit expression for $\mathcal{F}[{\bf X}]$ in terms of the source stress tensor. Let us write $\mathcal{F}[{\bf X}] = e^{i (S_{SG} + \chi[{\bf X}]})$, where $S_{SG}$ is a self-gravity term, and $\chi[{\bf X}] = \Delta[{\bf X}] + i\Gamma[{\bf X}]$ is the complex influence functional phase. Then at temperature $T=0$ we have \cite{gravDeco,oniga16}
\begin{align}
i\chi[{\bf X}] \;=\; &4\pi G  \int {d^3 q \over |{\bf q}|} \Pi^{\mu\nu\alpha\beta} \biggl(  \bigl[
T_{\mu\nu}^*(q)  T'_{\alpha\beta}(q) - c.c. \bigr] \nonumber \\
& \qquad\qquad\qquad\qquad - \delta  T_{\mu\nu}^*(q) \delta  T_{\alpha\beta}(q)   \biggr)
 \label{IFT}
\end{align}
in which $\delta T(q) \equiv T(q) - T'(q)$, ${\bf q}$ is a 3-momentum, and
$\Pi^{\mu\nu\alpha\beta} \equiv \sum_{\lambda} \epsilon^{\mu\nu}_\lambda \bigl[
\epsilon^{\alpha\beta}_{\lambda} \bigr]^\dagger $ is the transverse-traceless projector over graviton modes, with $\lambda = \pm$, so that $\chi[{\bf X}]$ is written in terms of the
transverse traceless part $h^{TT}$ of $h^{\mu\nu}$, projected out by $\Pi^{\mu\nu\alpha\beta}$.

The 2nd (purely real) term in (\ref{IFT}) is the ``decoherence functional'' $\Gamma[{\bf X}]$
\cite{decoF}; it completely describes the information loss as a functional of the pair of paths ${\bf X} = (X,X')$, and is thus the starting point of our analysis.

To see how a result like (\ref{IFT}) can be used in practise, consider the simple case where ${\cal S}$ is a particle, following a path with coordinate $X^{\mu}(s)$ and four-velocity $U^\mu = dX^\mu/ds$. Here $s$ is the usual proper time \cite{jackson}; one also defines \cite{jackson,suppInfo} the retarded time $t_r$ and the Bondi time $u = t-r$, where $r$ is the distance from the source at retarded time $t_r$ to the field point $x$.

The particle has stress tensor
\begin{equation}
T^{\mu\nu}(x) = m\int ds \, U^\mu(s) U^\nu(s) \delta^{(4)}(x - X(s) ),
 \label{eq:Stress}
\end{equation}
whose Fourier transform $T^{\mu\nu}(q) = \int d^4x \, e^{iq\cdot x} T^{\mu\nu}(x)$ is
\begin{equation}
 \label{eq:TofQ}
T^{\mu\nu}(q) = \frac{im}{\omega} \int ds \, e^{i\omega \hat l \cdot X(s)}  \partial_s
\left[\frac{{U}^{\mu}(s) {U}^{\nu}(s)}{\hat l\cdot U(s)}\right] ,
\end{equation}
where the 4-vector $q^\mu \equiv \omega\hat l^\mu \equiv \omega (1, \hat n)$, with $\hat n \cdot \hat n = 1$.

We begin by rewriting things in terms of the metric perturbation $h^{\mu\nu}(x)$, caused by
${\cal S}$. In harmonic gauge we write the trace-reversed $\bar{h}^{\mu\nu}$ in Li\'enard-Wiechert form as
\begin{equation}
\bar{h}^{\mu\nu}(u,r,\hat{n}) = 4mG\frac{U^\mu(t_r) U^\nu(t_r)}{|\vec x - \vec X(t_r)| \, \hat{l} \cdot U(t_r)}
\end{equation}
where we take the field point at a fixed Bondi time $u \equiv t - r$, and $\hat n$ is now the direction from the source to the field point. We then relate $\bar{h}^{\mu\nu}$ to $T^{\mu\nu}$ by taking its derivative along $u$ near $\scr{I}^+$, and then Fourier transforming:
\begin{align}
\partial_u\bar{h}^{\mu\nu}(\omega,\hat{n}) &=\; \int du \, e^{i\omega u} \partial_u
\lim_{r\rightarrow \infty} r \,\bar{h}^{\mu\nu}(u,r\hat{n}) \nonumber \\
&= 4mG \int ds \, e^{i\omega \hat l \cdot X(s)} \partial_s  \left[ \frac{U^\mu(s)
U^\nu(s)}{\hat{l} \cdot U(s)} \right]
 \label{eq:hDot}
\end{align}
where we use $u(X(s), \hat n) = \hat l(\hat n) \cdot X(s)$ to lowest order in $r^{-1}$ (see ref. \cite{suppInfo} for a more complete discussion).

Comparing (\ref{eq:TofQ}) and (\ref{eq:hDot}) we see that we have
\begin{equation}
 T^{\mu\nu}(\omega, \hat{n}) =  \frac{i}{4G\omega}   \int du \, e^{i\omega u} \partial_u  \lim_{r\rightarrow \infty} r \,\bar{h}^{\mu\nu}(u,r\hat{n}) 
 \label{T-q}
\end{equation}
for the stress tensor as a function of frequency $\omega$, on a sphere of radius $r$ centred on the source.

The phase space of gravitational radiation at null infinity can be parametrized by the Bondi news \cite{bondi} tensor (equivalently, the asymptotic shear), which we can write in the linearized regime as \cite{ash-bong}
\begin{equation}
N_{ab} \;=\;  \frac{1}{\sqrt{2}} \partial_u
\lim_{r\rightarrow \infty} r (h_{ab})^{TT}
 \label{Nab-h}
\end{equation}
where $a,b$ are indices of the angular coordinates perpendicular to the vector ${\bf r}$. The result for the influence function follows immediately from (\ref{IFT}) and (\ref{T-q}):
\begin{align}
i\chi[{\bf N}] \;=\;  \frac{1}{32\pi^2G}  \int &\frac{d\omega}{\omega} dS^2 \biggl( \bigl[
N_{ab}^*(\omega, \hat n) N'^{ab}(\omega, \hat n) - c.c. \bigr] \nonumber \\
  & - \delta N_{ab}^*(\omega, \hat n) \delta N^{ab}(\omega, \hat n)   \biggr)
 \label{IFT-N}
\end{align}
where ${\bf N} \equiv (N, N')$, and we have written $N_{ab}$ as a function of on-shell frequency
and direction $\hat{n}$ on the sphere $S$. Clearly, in view of the ``triangle'' connection between gravitational memory, asymptotic BMS charges, and ``soft factors''
\cite{strominger18,ashtekar18,strominger1,pasterski}, one can also expect to write the influence functional phase, evaluated on scattering processes, in terms of these quantities \cite{wilson18}.


\vspace{2mm}
{\bf (ii) Decoherence and Asymptotic States:} Our results in linearized
gravity are consistent with others derived for soft gravitons
\cite{strominger18,ashtekar18,strominger1,pasterski,wilson18,carney}. However, any discussion of, eg., black holes, would require that we go beyond linearized gravity, to the full non-linear theory.

To make contact with the full theory, we derive an expression for the decoherence functional in terms of ``asymptotic states'' $|| \sigma \rangle\rangle$, which can be thought of as coherent quantum states of the metric, defined on $\scr I^+$, and centred on a particular classical configuration of the asymptotic shear $\sigma$ \cite{ashtekar81}. These states, which represent the true asymptotic radiative degrees of freedom, are not tied to linearized gravity, although they reduce to ordinary gravitons in the linearized regime. To define them, one introduces a symplectic form $\Omega(\sigma, \sigma')$ for gravitational excitations on a Cauchy slice $\Sigma$; then, letting $\Sigma \rightarrow \scr{I}^+$, one
has \cite{ash-M80}
\be
\Omega(\sigma, \sigma') = \frac{1}{16\pi G} \int d \scr{I}^+ \bigl( \sigma^{ab} {\cal N}'_{ab}  -
{\cal N}^{ab}\sigma'_{ab}  \bigr) .
\ee
where ${\cal N}_{ab}$ is the Bondi news tensor for full General Relativity (only reducing to $N_{ab}$ in (\ref{Nab-h}) in the linearized limit). It is given in terms of the shear tensor $\sigma_{ab}$ by
\begin{equation}
{\cal N}_{ab} \;=\; 2 \partial_u \sigma_{ab}; \qquad \;\; {\cal N}^{\prime}_{ab} \;=\; 2 \partial_u \sigma^{\prime}_{ab}
 \label{Nab-GR}
\end{equation}

One also introduces a complex structure $J$, which acts on the fields as $J \cdot \sigma_{ab} = i \sigma^+_{ab} - i\sigma^-_{ab}$, where $\pm$ denote the positive/negative frequency parts.

We can then define quantized operators $\hat{\cal N}[f]$ corresponding to the Bondi news smeared out using some test function $f$ thought of as a particular classical choice of $\sigma$. These are
\be \label{smearedN}
\hat{\cal N}[f] \equiv -\frac{1}{8\pi G} \int d \scr{I}^+ \hat{\cal N}^{ab}f_{ab}
\ee
with commutator
\be \label{eq:commutator}
\bigl[ \hat{\cal N}[f], \hat{\cal N}[f']   \bigr] = i\Omega(f,f') \mathbb{I} .
\ee

The coherent states are then created by exponentiating these smeared news operators and acting on the vacuum:
\be \label{eq:coherent}
||\sigma\rangle\rangle \equiv e^{i \hat{\cal N}[\sigma]} |0\rangle  \equiv \hat W[\sigma] |0\rangle
\ee
where $|0\rangle$ represents a coherent state vacuum peaked at a particular classical spacetime with no radiation. Here, we will assume this vacuum to be a semi-classical state with asymptotic Minkowski metric. The $\hat W$ operators have the vacuum expectation value
\be \label{eq:VEV}
\langle0|\hat W[\sigma]|0\rangle = e^{-\frac{1}{2}\Omega(\sigma, J\cdot \sigma)}.
\ee
so that, from \eqref{eq:commutator} and \eqref{eq:VEV}, the overlap between two coherent states is
\be
\begin{split}
\langle\langle\sigma'|\sigma\rangle\rangle &= \langle0|\hat W^\dagger[\sigma'] \hat
W[\sigma]|0\rangle  \\
&= e^{  -\frac{1}{2}\Omega(\sigma-\sigma', J\cdot (\sigma-\sigma')) +
\frac{i}{2}\Omega(\sigma',\sigma) }
\end{split}
\ee
or, rewriting the exponential in momentum space, we have an overlap
\begin{align}
\exp \biggl\{   & \frac{1}{32 \pi^2 G}   \int_0^\infty \frac{d\omega }{\omega} dS^2 \biggl(
\bigl[  {\cal N}^*_{ab}(\omega,\hat n) {\cal N}'^{ab}(\omega, \hat n) - c.c. \bigr] \nonumber \\
 & \qquad -    \delta {\cal N}^*_{ab}(\omega, \hat n) \delta {\cal N}^{ab}(\omega, \hat n)    \biggr)   \biggr\}
 \label{eq:overlap}
\end{align}
which reduces to (\ref{IFT-N}) in the linearized limit.

We now observe that if the asymptotic states $\sigma, \sigma'$ result from the pairs of paths (forward and reversed) in the evolution of the reduced density matrix of some matter system, we can then immediately write the influence functional in the very simple form
\begin{equation}
 \label{result}
{\cal F}[\mbox{\boldmath $\sigma$[\bf X]}] = e^{i\bf{S}_{SG}[\bf X]}
\langle\langle\sigma'|\sigma\rangle\rangle
\end{equation}
where, as before, $\bf{S}_{SG}[\bf X]$ is a ``self-gravity'' term, and $\Delta[\boldmath \sigma]$ and $\Gamma[\boldmath \sigma]$ are now written as functionals of $\mbox{\boldmath $\sigma$} \equiv (\sigma, \sigma')$. The imaginary part $\Gamma[\mbox{\boldmath $\sigma$}]$ (the decoherence functional) is then just the functional
\begin{equation}
\Gamma[\mbox{\boldmath $\sigma$}] = - \ln | \langle\langle\sigma'|\sigma\rangle\rangle |
 \label{Gamma-S}
\end{equation}
of the configurations $\sigma_{ab}(x), \sigma_{ab}'(x)$ at $\scr{I}^+$.

It is important to understand that we can only write the decoherence functional explicitly in terms of the source variables $T,T'$ in the linearized limit (see (\ref{IFT-N})). In the full non-linear case, the states $\boldmath \sigma = (\sigma, \sigma')$ are generated not only by the matter sources, but also by the vacuum curvature sources accompanying them, and there is no generally known map between the matter source and field configuration for the full Einstein equations \cite{newman95}.

We note that quite generally, even if we do not know the connection between the time evolution of ${T}_{\mu\nu}$ and the news function, given an asymptotically flat space-time we can still write a density matrix for the total matter-plus-gravity system at late times. Schematically, we expect this to have the form $\mbox{\boldmath $\rho$} = |\Psi_f\rangle \langle \Psi_f |$, where the ``out state'' is
\begin{equation}
 \label{nonlin}
|\Psi_f\rangle \;=\; \sum_i a_i |M_i\rangle ||\aleph_i \rangle \rangle
\end{equation}
in which the $|M_i\rangle$ are states of the non-gravitational degrees of freedom, and the $||\aleph_i \rangle \rangle $ are the out states of the gravitational degrees of freedom. Note that the $||\aleph_i \rangle \rangle $ will not in general be single coherent states $||\sigma_i \rangle\rangle$, but rather a superposition of them, ie., $||\aleph_i \rangle \rangle  = \sum_{\alpha} b_{i \alpha} || \sigma_{\alpha}\rangle\rangle$. Constructing the reduced density matrix $\bar{\boldsymbol \rho} = Tr_g \mbox{\boldmath $\rho$}$, it is easy to see that the interference terms will decay as $|\langle\langle \aleph_i | \aleph_j \rangle\rangle|$, and we can define an effective ``decoherence functional'' $\Gamma_{ij}  = - \ln|\langle\langle \aleph_i | \aleph_j \rangle\rangle|$. We regard $\Gamma_{ij}$ as a functional in the sense that it depends on entire histories of the matter, as the asymptotic states are defined on all of $\scr I^+$. Regarding the full theory, what we have done is simply point out that these asymptotic states - and associated shear variables - are the natural ones for describing time-dependent information loss in quantum systems. Extension of the calculations to extended massive bodies, and to non-linear cases in which $T_{\mu\nu}$ involves contributions from the metric itself, will be the subject of future work (even in the linearized regime, this formalism greatly simplifies calculations in which there is a sensible semiclassical limit).


\vspace{2mm}

{\bf (iii) Two-Path Experiment:} Consider a setup where the matter paths fall into two distinct classes \cite{2class}, which we label $1$ and $2$; this is a very simple example of a system possessing ``internal states" which couple to $g^{\mu\nu}(x)$. A classic example is a ``2-slit" system \cite{baym09}, where a mass $m_o$ passes through a slit assembly of mass $M_A$ through two (assumed equally likely) classes of path, associated with one or other of the slits; for consistency we need to follow the dynamics of both masses \cite{2mass,suppInfo}. Then, once the two masses have interacted and separated, and the disturbance in $g^{\mu\nu}$ has propagated off towards ${\scr I}^+$,
the probability ${\cal P}(z_f)$ of arrival of the mass $m_o$ at a final specific point $z_f$ on the screen is then usually written ${\cal P}(z_f) = {\cal P}_1(z_f) + {\cal P}_2(z_f) + {\cal P}_{12}(z_f)$, where ${\cal P}_{12}$ is the usual interference term.

If we ignore the coupling of these masses to other non-gravitational fields (which will excite photons, phonons, etc. \cite{2slitP}), then we can easily analyze this system.

Consider first the linearized case. Defining $\bar{T}^{\mu\nu} = T_s^{\mu\nu} + T_A^{\mu\nu}$, the sum of the stress tensors for the 2 masses, we write $\delta_{12} \bar{T}^{\mu\nu} = \bar{T}^{\mu\nu}_1 - (\bar{T}_2^{\prime})^{\mu\nu}$ as the difference between $\bar{T}$ on an outgoing path $1$ and $\bar{T}'$ on a return path $2$, and the analogous difference of news tensors as $\delta_{12} \bar{N}_{ab}$. Then, by a straightforward application of \eqref{result}, along with a semiclassical approximation restricting the path integral to the classical paths 1 and 2, we find that ${\cal P}_{12}(z_f) = P^o_{12}(z_f)| \langle\langle\sigma_1|\sigma_2\rangle\rangle |$, where $P^o_{12}(z_f)$ incorporates all terms apart from the decoherence, and the decoherence rate $\Gamma_{12}$ is then
\begin{align}
\Gamma_{12} \;&=\; - \ln | \langle\langle\sigma_1|\sigma_2\rangle\rangle | \\
 & = \frac{1}{32\pi^2G}  \int \frac{d\omega}{\omega} dS^2 \; \delta_{12} \bar{N}_{ab}^*(\omega,
 \hat n) \delta_{12} \bar{N}^{ab}(\omega, \hat n)   \nonumber \\
& = 4\pi G  \int {d^3 q \over |{\bf q}|} \Pi^{\mu\nu\alpha\beta}
 \delta_{12} \bar{T}_{\mu\nu}^*(q) \delta_{12} \bar{T}_{\alpha\beta}(q)  \nonumber
 \label{linG-2p}
\end{align}
in the linearized limit. The asymptotic states thus allow us to quickly recover the original influence functional.

If we now go to the general non-linear case, then using (\ref{nonlin}) for each class of matter paths $i=1,2$ associated with slits $1$ and $2$, we can write
\begin{equation}
\Gamma_{12} \;=\; - \ln \langle\langle \aleph_2 | \aleph_1 \rangle\rangle = - \ln \sum_{\alpha, \beta} b^*_{2\alpha} b_{1\beta} \langle\langle \sigma_{\alpha} | \sigma_{\beta}\rangle\rangle
  \label{G-aleph}
\end{equation}
where $||\aleph_i \rangle \rangle  = \sum_{\alpha} b_{i \alpha} || \sigma_{\alpha}\rangle\rangle$ as before. To do quantitative calculations we would need of course to know the coefficients $\{ b_{i \alpha} \}$.


\vspace{2mm}

{\bf (iv) Entanglement Measures}: Let us now conclude by considering various additional ways of characterizing the information loss from the 2-path system caused by the entanglement with $g^{\mu\nu}(x)$; these are particularly simple when expressed in terms of these overlaps. In particular one has

(i) the {\it purity} of a reduced density matrix $\bar{\rho}$, defined as $\scr{P} \equiv \mathrm{Tr}\bigl[ \bar{\rho}^2 \bigr]$. This is immediately given here by
\be
\scr{P}(t) = \frac{1}{2}\biggl(  1 + | \langle\langle \aleph_2 | \aleph_1 \rangle\rangle  |^2 \biggr).
\ee

(ii) The {\it von Neumann entanglement entropy}, $\scr{S}(\bar \rho) \equiv - \mathrm{Tr}\bigl[ \bar \rho \ln\bar\rho \bigr]$, which is found to be
\be
\begin{split}
\scr{S}(t) = \ln2 - \frac{1}{2} \bigl[ &(1+ | \langle\langle \aleph_2 | \aleph_1 \rangle\rangle  |) \ln (1+ | \langle\langle \aleph_2 | \aleph_1 \rangle\rangle  |) \\
+ \; & (1- | \langle\langle \aleph_2 | \aleph_1 \rangle\rangle  |) \ln (1- | \langle\langle \aleph_2 | \aleph_1 \rangle\rangle  |)     \bigr]
\end{split}
\ee

(iii) Finally, the {\it fidelity} of our 2-path system, with respect to some pure reference state $\rho^0 $; this is defined as $\scr{F}(t) \equiv \mathrm{Tr} \bigl[  \rho^0 \bar{\rho}(t) \bigr]$. A natural choice of reference state is that obtained by unitary time evolution of the initial state in the absence of radiation, so that  $\rho^0(t) = \hat{U}^0(t) \rho(0) \hat{U}^{0 \dagger}(t) $, in which $\hat{U}^0(t) $ is the free time evolution operator for the mass $M$. The fidelity then defines the overlap between these two states, and one gets
\begin{equation}
\scr{F}(t) = \frac{1}{2} \bigl[  1 + \mathfrak{Re}  \langle\langle \aleph_2 | \aleph_1 \rangle\rangle   \bigr] .
\end{equation}

Which of these entanglement measures is most useful depends on the circumstances \cite{entM}; the point we wish to emphasize is that we can give simple closed expressions for them which still encapsulate all of the relevant time dependence. Finally, we note that in the linearized regime, the above expressions for these information theoretic quantities have exactly the same form, with $\langle\langle \aleph_2 | \aleph_1 \rangle\rangle$ replaced by $\langle\langle \sigma_2 | \sigma_1 \rangle\rangle$.


\vspace{2mm}

{\bf (v) Conclusion}: In linearized gravity one can discuss information loss in scattering calculations in terms of an infrared `triangular' relation between BMS charges, soft factors, and linearized gravitational memory. Outside the framework of scattering calculations, and for the full non-linear theory, we have shown that the natural formulation of decoherence and information loss involves overlap integrals between asymptotic Ashtekar states, or between superpositions of these. This is not only of theoretical interest - it also yields practically useful results, as we see from our 2-state example.

This research was supported by the National Sciences and Engineering Research Council of Canada.

\clearpage 
\onecolumngrid

{\centerline{\bf SUPPLEMENTAL INFORMATION}} 
\vspace{3mm}

In this supplementary information we give a few more details on the geometrical notions being used in the paper. We first define the coordinates we use for the description of particle motion and for quantities like the Bondi news function, in classical relativity; and we show how these relate to some other commonly used conventions. We then briefly describe the 2-slit geometry used in the paper. 

\vspace{3mm}
{\bf 1. NULL COORDINATES:}

\begin{figure}[h]
\includegraphics[width=12cm]{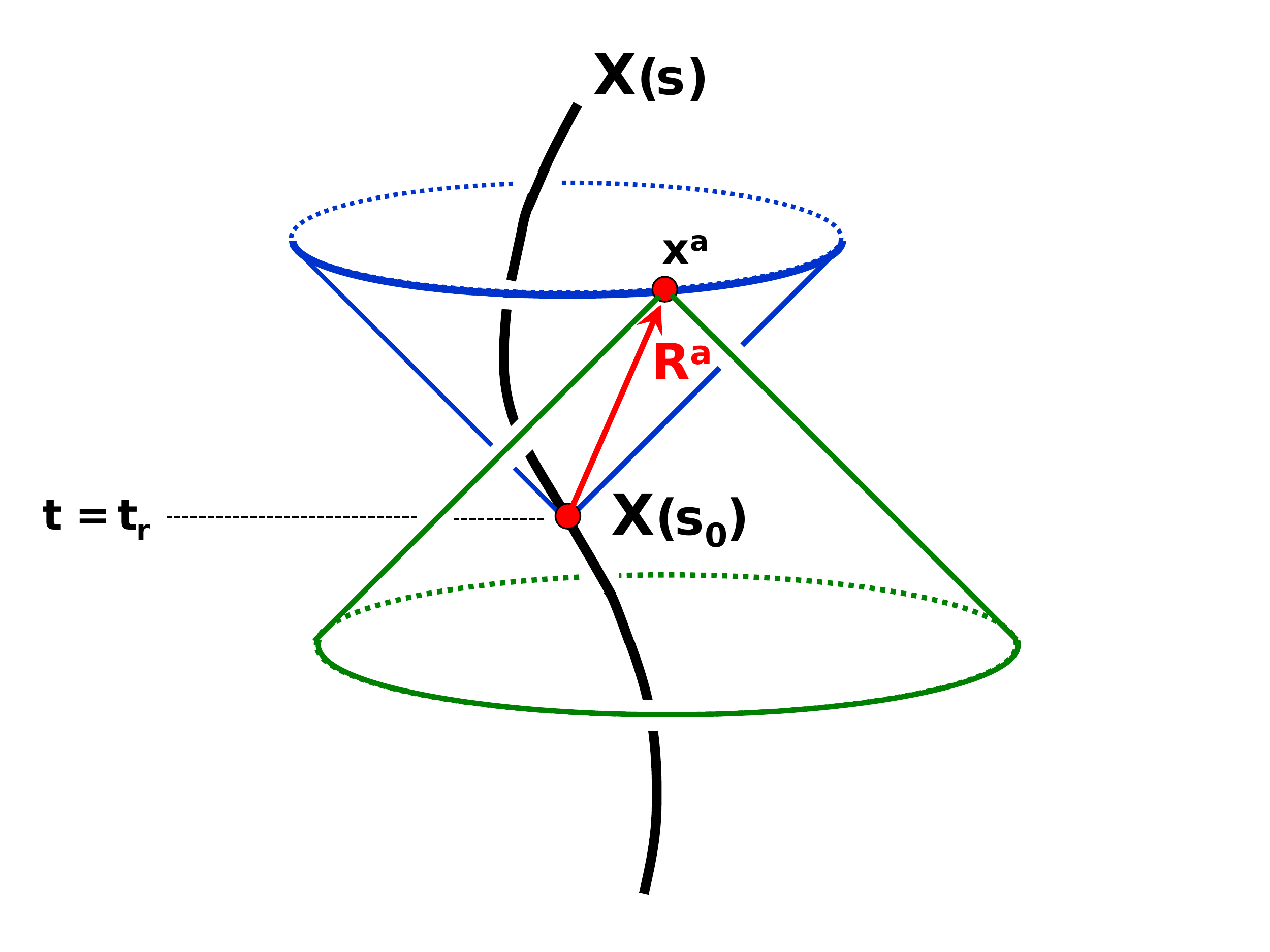}
\caption{ Spacetime diagram illustrating the terms in the text. A particle follows the trajectory $X(s)$ as a function of proper time $s$. When $s=s_o$, the coordinate time $t=t_r$, where $t_r$ is the retarded time. The ``field point'' at $x = x^a$ is on the future light cone of $X(s_o)$, ie., $X(s_o)$ is the point where the past light cone of $x$ intersects the particle trajectory; they are separated by the null vector $R^a = x^a - X^a(s_o)$ lying on the intersection of the two cones. The time axis is vertical, and the horizontal plane represents two of the 3 spatial dimensions. }
 \label{fig:colby-F1}
\end{figure}

The conventions used in the main text, for the discussion of particle dynamics, are most easily explained using Fig. \ref{fig:colby-F1}. We will be discriminating between 4 different time variables, these being $s$ (the particle proper time), $t$ (the coordinate time), $u$ (the coordinate Bondi time), and $t_r$ (the retarded time).


We start by considering a point in flat spacetime $x = x^a$, parameterized by the spherical coordinates $(t, r, \theta, \phi)$. A field disturbance field at the point $x$ must be influenced, via the gravitational field, by the state of some source at a point intersecting the past light cone of $x$. For a timelike trajectory of the source, this happens at exactly one point. The time coordinate $t$ at this point is the retarded time $t_r$. To define the relation between the two, we pick some value of the proper time $s = s_0$, and define the retarded time as $t_r = X^0(s_0)$, where $X^a(s_0)$ is chosen to be the point of intersection of $X(s)$, the source trajectory,  with the past light cone of $x$. Then, reparameterizing $X(s) \rightarrow X(t)$, so that $X^0(t_r) = t_r$, we enforce the condition that the point $X(t_r)$ is null separated from $x$; because we are interested in the retarded potential, we assume $t > t_r$, so this null condition becomes
\begin{equation}
t - t_r = |\vec x - \vec X(t_r)|
 \label{t_r}
\end{equation}

Now let us take the source to be a particle of mass $m$, following an arbitrary (timelike) trajectory $X(s)$, where $s$ is now the particle proper time. We write $R^a \equiv x^a - X^a(t_r)$ for the separation 4-vector between the source and the spacetime point at $x$, and $U^a \equiv dX^a/ds$ for the particle four-velocity. This separation vector $R^a$ is null, so we rewrite it as $R^a \equiv R \hat l^a = R (1, \hat n)$, where $\hat n$ is a spatial unit vector pointing from $\vec X(t_r)$ to $\vec x$, and $R \equiv |\vec x - \vec X(t_r)|$.

The linearized equations describing the perturbation generated by the particle then have, in harmonic gauge, the usual retarded Li\'enard-Wiechert (LW) solution (obtained in complete analogy with the electromagnetic case \cite{Jackson})
\begin{equation}
 \label{eq:LW}
\bar h^{ab}(x) \equiv \bar h^{ab}(t, r, \theta, \phi) = 4m G \frac{U^a(t_r) U^b(t_r)}{R\cdot U(t_r)} \;\; \equiv \;\; 4m G \frac{U^a(t_r) U^b(t_r)}{R \hat l\cdot U(t_r)}
\end{equation}
as given in the text.

We are interested in looking at the leading order LW field near future null infinity $\scr I^+$. A convenient choice of coordinates is the Bondi coordinate system, using the null time variable $u \equiv t - r$ (the Bondi time). Note that changing $u$, while keeping $r, \theta, \phi$ fixed, traces out ingoing null rays; whereas changing $r$, while keeping $u, \theta, \phi$ fixed, traces out outgoing null rays. In Bondi coordinates, $\scr I^+$ is reached by taking $r \rightarrow \infty$ while holding $u$ fixed. To isolate the leading behavior of the field near null infinity, we want to compute: $\lim_{r\rightarrow \infty} r \bar h^{ab}(u,r,\theta, \phi)$.

To pick out the large-$r$ behavior, we make the usual multipole expansion of $1/|\vec x - \vec X(t_r)|$.  The approximation that $t_r \approx u = t-r$ is commonly made at this point, ie., we assume that since $R$ is very large, the time that radiation takes to propagate from the source to $x$ is approximately the same as it takes to propagate from the origin to $x$. To properly capture the leading behavior, we must go beyond this assumption; we thus write
\begin{equation}
t_r = t - |\vec x - \vec X(t_r)| \approx t-  r + \hat n \cdot \vec X(t_r) + \mathcal{O}(r^{-1})
 \label{t-Ret}
\end{equation}
so that, in terms of $u$, we have
\begin{equation}
 \label{light_cone_cond}
u = t_r -  \hat n \cdot \vec X(t_r) \equiv X^a(t_r) \hat l_a
\end{equation}

Going back to \eqref{eq:LW}, We can now state the limit to null infinity. From (\ref{eq:LW}) we have
\begin{equation}
\bar h^{ab}(u, r, \theta, \phi) \;=\;  4m G \frac{U^a(t_r) U^b(t_r)}{R \hat l\cdot U(t_r)}  \;\;\approx \;\;   4m G \frac{U^a(t_r) U^b(t_r)}{r \hat l\cdot U(t_r)} + \mathcal{O}(r^{-2})
\end{equation}
using $t_r \approx u + \vec X(t_r) \cdot \hat n$. Then the Fourier transform of the Bondi time derivative near null infinity is
\begin{align}
 \int du \, e^{i\omega u} \partial_u
\lim_{r\rightarrow \infty} r \,\bar{h}^{ab}(u,r, \theta, \phi)  &= \;4m G \int du \, e^{i\omega \hat l \cdot X(t_r) } \partial_u \biggl[ \frac{U^a(t_r) U^b(t_r)}{\hat l\cdot U(t_r)}    \biggr] \nonumber \\
&= \; 4m G \int dt_r \, e^{i\omega \hat l \cdot X(t_r) } \partial_{t_r} \biggl[ \frac{U^a(t_r) U^b(t_r)}{\hat l\cdot U(t_r)}    \biggr]
\end{align}
using $\partial_u = (dt_r/du)\partial_{t_r}$.  If instead we parametrize the result in terms of proper time, then using the same trick we get
\begin{equation}
 \label{propertime}
 \int du \, e^{i\omega u} \partial_u
\lim_{r\rightarrow \infty} r \,\bar{h}^{ab}(u,r, \theta, \phi) \;\;=\;\; 4m G \int ds \, e^{i\omega \hat l \cdot X(s) } \partial_{s} \biggl[ \frac{U^a(s) U^b(s)}{\hat l\cdot U(s)}    \biggr]
\end{equation}
which is the result given in eq. (7) of the main text.


\begin{figure}
\includegraphics[width=12cm]{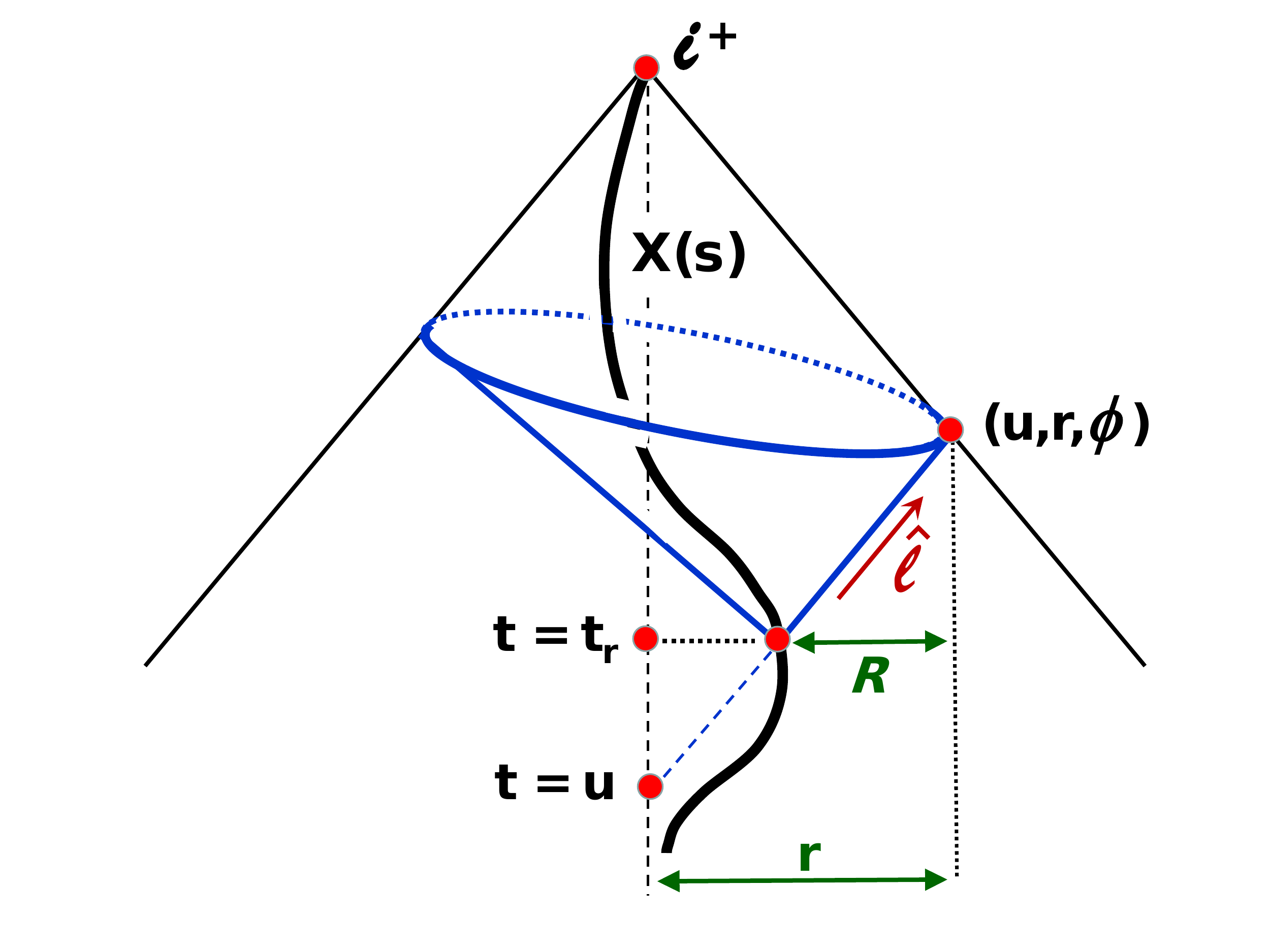}
\caption{ The relationship between the different coordinates used. The particle follows the path $X(s)$ as a function of proper time $s$; we show it at coordinate time $t=t_r$, where $t_r$ is the retarded time, so that the Bondi time $u$ is then $u = t_r -  \hat n \cdot \vec X(t_r) \equiv X^a(t_r) \hat{\ell}_a$ (compare eq. (\ref{light_cone_cond})), where $\hat{\ell}$ is a null vector, and we see the angular dependence of the Bondi time corresponding to the arrival of null radiation at $\scr I^+$. The vertical hatched line indicates the origin $r=0$, which terminates at the point $i^+$ on $\scr I^+$.}
 \label{fig:colby-f2}
\end{figure}


It is essential to keep the correction in (\ref{light_cone_cond}) to the retarded time  - it is equivalent to taking into account the angular dependence of the intersection of the particle's future light cone with $\scr I^+$, and it makes sure we keep track of all the phase information in the stress tensor, via
\begin{equation}
 T^{ab}(\omega, \hat{n}) =  \frac{i}{4G\omega}   \int du \, e^{i\omega u} \partial_u  \lim_{r\rightarrow \infty} r \,\bar{h}^{ab}(u,r,\theta, \phi) .
 \label{T-q}
\end{equation}

We see this in Fig. 2, which illustrates the difference between the various time coordinates. The Bondi time $u$ equals the retarded time $t_r$ if $\vec X(t_r) = 0$, ie., the particle is at the origin. Otherwise, the correction is the particle displacement $\hat n \cdot \vec X(t_r)$ in the $\hat n$ direction: if the particle is closer to $\scr I^+$ in this direction, radiation takes less time to arrive.

We note that, comparing eq. (\ref{propertime}) to the Fourier transform of the particle stress tensor
\begin{equation}
 \label{eq:TofQ}
T^{ab}(q) = \frac{im}{\omega} \int ds \, e^{i\omega \hat l \cdot X(s)}  \partial_s
\left[\frac{{U}^{a}(s) {U}^{b}(s)}{\hat l\cdot U(s)}\right]
\end{equation}
shows us that lining up the lightcones properly is extremely important!  Without understanding the light cone cuts properly, we lose phase information in $T^{ab}$. That these cuts contain information is not new. \cite{newman83}

\vspace{6mm}

{\bf 1. 2-PATH EXPERIMENT:}

There are many ways to set up a 2-path experiment of the kind discussed in the text. To get an idea of how this works, it is useful to consider a concrete geometry. In Fig 2 we show a simple 2-slit system - although this would not be practicable for any real experiments, it suffices to show what is involved. For a somewhat different analysis, see Baym and Ozawa \cite{baym09}.


\begin{figure}
\includegraphics[width=10cm]{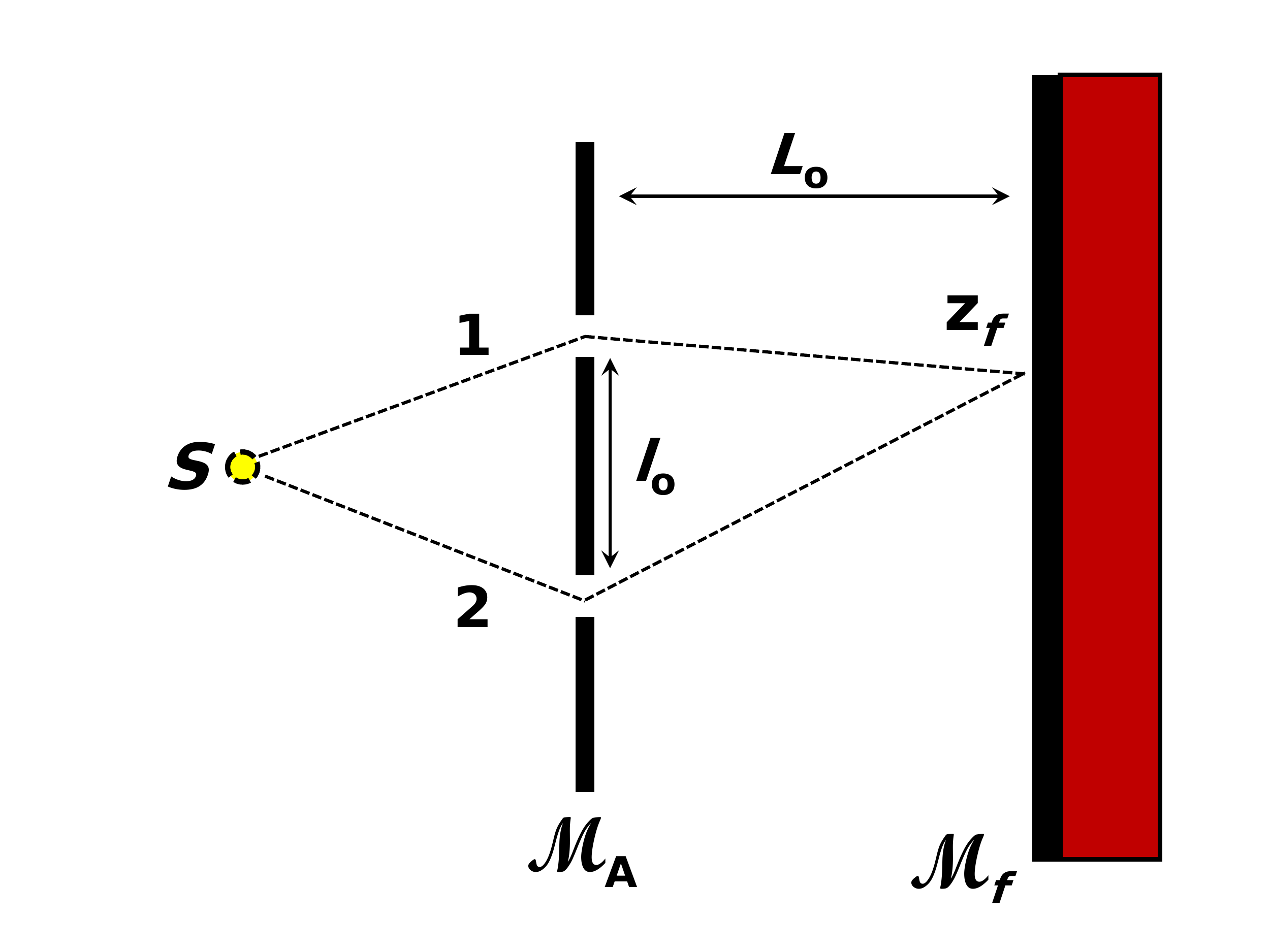}
\caption{ The 2-slit system to be studied here. The experimental system ${\cal S}$ is a moving mass $m_o$, which passes along one of two paths (labeled $1$ and $2$) through a 2-slit system ${\cal M}_A$ of mass $M_A$. The slits are a distance $l_o$ apart. The moving mass then passes on to a screen ${\cal M}_f$ of mass $\gg M_A$, at a distance $L_o$ from ${\cal M}_A$, arriving at $z$-coordinate $z_f$ on the screen. }
 \label{fig:colby-f2}
\end{figure}


In this setup, the system ${\cal S}$ of interest is a mass $m_o$ which passes through either slit 1 or slit 2, with typical paths shown schematically in the figure. It is essential to take account of the dynamics of the slit system ${\cal M}_A$, itself of mass $M_A$, to get the correct coupling of the combined system to $g^{\mu\nu}(x)$. If the mass $m_o$ is moving at a velocity $v_o$, then for the setup shown, we expect a recoil of the slit system ${\cal M}_A$, roughly along the vertical $\hat{z}$-axis, with velocity
\begin{equation}
\Delta v_A \sim {m_o \over M_A} v_o {l_o \over L_o}
 \label{recoil}
\end{equation}
where, as shown in Fig. 3, $l_o$ is the distance between the slits, and $L_o$ is the distance between the slit system ${\cal M}_A$ and the final screen system ${\cal M}_f$. In principle we should even take account of the dynamics of ${\cal M}_f$, but with no loss of generality we will assume that the mass of the screen system is so large, and its recoil so small, that its dynamic coupling to $g^{\mu\nu}(x)$ can be ignored.

Under these circumstances, the behaviour of the radiative modes of the gravitational field is determined in linearized gravity by the interaction term ${\cal L}_{int} = -\tfrac{1}{2} \int d^4 x h^{\mu\nu}(x) \bar{T}_{\mu\nu}(x)$ between the linearized gravitational field $h^{\mu\nu}(x)$ and the stress-energy tensor $\bar{T}_{\mu\nu}(x) = T_s^{\mu\nu} + T_A^{\mu\nu}$ for the combined system ${\cal S} + {\cal M}_A$; once we know the detailed dynamics for $\bar{T}_{\mu\nu}(x)$, we can calculate the behaviour of $h^{\mu\nu}(x)$ and of the Bondi news function as described in the paper.

One can also estimate the order of magnitude of any effects, by noting that they will depend in the usual way on the 3rd time derivative $\dddot{Q}$ of the quadrupole moment $Q(t)$ of the combined system ${\cal S} + {\cal M}_A$. If we know that the interaction between ${\cal S}$ and ${\cal M}_A$ takes place over a timescale $\tau_o$, and involves a displacement of the mass $m_o$ over a length scale $r_o$, then, provided $M_A \gg m_o$, we have $\dddot{Q} \sim m_o r_o l_o/\tau_o^3$. This then allows us to make simple estimates for the decoherence and information loss from the total system, without having to calculate the detailed behaviour of $\bar{T}_{\mu\nu}(x)$.

We note that Fig. 3 is schematic in the sense that in reality, the {\it single} path shown passing through slit $1$ actually represents the set of {\it all} paths leaving slit $1$ on their way to the screen (this shorthand way of representing entire classes of path appears to have originated in the celebrated description of 2-path experiments given by Feynman \cite{feynman65}). One can make an unambiguous separation between different classes of paths by introducing surfaces of ``final crossing'' of the paths - see eg., Auerbach and Kivelson \cite{auerbach84}. In the present case the set of all paths through slit $1$ is then simply defined as the set of all paths originating at the source $S$, and terminating on the screen system ${\cal M}_f$, whose last passage through the slit system ${\cal M}_A$ happens to be through slit $1$.



\begin{thebibliography}{99}
\bibitem{2mass}  For a correct calculation of the gravitational disturbance coming from the source $T_{\mu\nu}$, we must keep both masses $m_s$ and $M_A$, so that $T_{\mu\nu}$ is conserved.
\end{thebibliography}
\end{document}